\documentclass[12pt]{article}
\usepackage{graphicx}
\usepackage{epsfig}
\title{\bf Long Range Electromagnetic Effects involving
Neutral Systems and Effective Field Theory}
\author{ Barry R. Holstein\\
Department of Physics-LGRT,\\
University of Massachusetts,\\
Amherst, MA 01003}
\begin{document}
\begin{titlepage}
\maketitle
\begin{abstract}
We analyze the electromagnetic scattering of massive particles with
and without spin wherein one particle (or both) is electrically
neutral.  Using the techniques of effective field theory, we isolate
the leading long distance effects, both classical and quantum
mechanical. For spinless systems results are identical to those
obtained earlier via more elaborate dispersive methods.  However, we
also find new results if either or both particles carry spin.
\end{abstract}
\end{titlepage}

\section{Introduction}

There has been a good deal of recent interest in higher order
corrections to electromagnetic scattering.  In particular the
one-photon-exchange approximation, which has traditionally been used
to analyze electron scattering has been shown to be inadequate when
applied to the problem of isolating nucleon form factors via
Rosenbluth separation---inclusion of two-photon-exchange
contributions has been found to be essential in resolving small
discrepancies with the values of these same form factors as obtained
from spin correlation measurements\cite{ros}. A second arena where
two-photon-exchange effects are needed is the in the analysis of
transverse polarization asymmetry measurements in electron
scattering.  Such quantities vanish in the one-photon-exchange
approximation meaning that the sizable effects found experimentally
must arise from two-photon effects\cite{tpa}.

Much has been written about such higher order photon processes and a
number of groups have undertaken precision calculation of such
effects\cite{the}.  It is not our purpose here to attempt such
detailed calculations of charged particle interactions or to
confront experimental data. Rather our goal is to use the methods of
effective field theory (EFT) in order to analyze the very longest
range (smallest momentum transfer) contributions to the
electromagnetic scattering process when one or both of the
scattering particles are neutral. These long range components are
associated with pieces of the scattering amplitude which are
nonanalytic (and singular) in the momentum transfer. Some of these
corrections are classical ($\hbar$-independent) and behave as
$1/\sqrt{-q^2},\sqrt{-q^2},etc.$ while others are quantum mechanical
($\hbar$-dependent) and behave as $\log -q^2,q^2\log-q^2,etc.$,
where $q^2$ is the momentum transfer\cite{jdh}.  In the case of two
spinless charged particles the lowest order interaction, which
arises from one-photon exchange, is the simple Coulomb interaction,
which behaves as $\alpha/r$, where $\alpha=e^2/4\pi$ is the fine
structure constant.  The contribution to this charged scattering
process from two-photon exchange is a problem addressed nearly two
decades ago by Feinberg and Sucher using dispersive
methods\cite{fs}.  Even earlier Iwasaki had studied the classical
piece of this problem using standard noncovariant perturbation
theory\cite{iwa}.  Recently we reexamined this problem, using the
methods of effective field theory (EFT)\cite{hr}.  Results for
spinless scattering were found to agree with those of \cite{fs} and
\cite{iwa}, but the use of EFT methods permitted the extraction of
new and interesting spin-dependent structure.

Our goal in the present note is to extend these considerations to
the case of the electromagnetic scattering of two nonzero mass
particles, at least one of which is neutral.  In this case there
exists no lowest order Coulomb potential and the leading
contribution arises from two-photon exchange.  The interaction of
two spinless systems was considered long ago by Casimir and
Polder\cite{cp} and by Feinberg and Sucher\cite{fs1} in the
neutral-neutral case and by Bernabeu and Tarrach\cite{bt} and by
Feinberg and Sucher in the case of the interaction of a neutral and
a charged particle\cite{fs2}. The first of these calculations was
performed using noncovariant fourth order perturbation theory, while
the latter evaluations were done using dispersive methods. In the
present paper we reanalyze these problem using EFT techniques.  The
basic idea is to calculate the infrared singular components of the
two-photon-exchange diagrams, since such terms give rise to the
longest order interactions in coordinate space.  In the case of
spinless scattering, we will reproduce the results of previous
authors\cite{fs1,fs2,bt}.  However, the use of EFT methods allows
the straightforward extraction of the new and interesting structure
which arise if either or both particles carry spin.

In the next section we study the interaction of two neutral
particles, while in the following chapter we look at the situation
when one of these particles is charged.  We present a brief
summary in a concluding section.

\section{Neutral-Neutral Scattering}

The electromagnetic interaction of two neutral systems having
separation $r$, the so-called Van der Waals force, was considered
long ago by London\cite{lo}, who gave a simple form for the
interaction potential in terms of the electric polarizabilities of
the two systems---
\begin{equation}
V_{vdW}(r)\sim -{\alpha_E^a\alpha_E^b\omega_0\over 4\pi r^6}
\end{equation}
where $\omega_0$ is a typical excitation energy.  The form of the
vanderWaals potential can be understood in terms of the energy of
the dipole moment of "atom" $b$ ($d_b=-ex_b$) in the electric field
created by the dipole moment of "atom" $a$---
\begin{equation}
{\cal H}_1\sim-d_bE_b(d_a)= ex_b\times {-ex_a\over 4\pi
r^3}=-{e^2x_ax_b\over 4\pi r^3}
\end{equation}
Of course, $<x_a>=<x_b>=0$, {\it i.e.}, there exists no average
dipole moment, so this energy change vanishes in first order
perturbation theory
$${\Delta E}_1=<\psi_0|{\cal H}_1|\psi_0>=0.$$
However, there {\it is} a shift at second order since at any given
instant of time there exists an {\it instantaneous} dipole moment in
atom $a$ say.  The corresponding electric field from atom $a$ at the
position of atom $b$---$E_a(R)$---generates a correlated electric
dipole moment due to its electric polarizability---
\begin{equation}
d_b=4\pi \alpha_E^b E_a(R)=4\pi\alpha_E^b {ex_a\over 4\pi r^3}
\end{equation}
The electric field generated by {\it this} electric dipole moment
then acts back on the original atom, yielding an energy
\begin{equation}
\Delta E_{vdw} \sim - d_aE_b(r)=-{e^2x_a^2\alpha_E^b\over 4\pi r^6}
\end{equation}
which is the Van der Waals interaction.  What makes this work,
then, is the point that one can use the {\it instantaneous}
position of one atom to provide an action at a distance
correlation with a second atom in the vicinity.  Finally, we note
that the electric polarizability itself can be extracted by
calculating the shift in energy of the atom in the presence of an
external electric field $E_0$ in second order perturbation theory
\begin{equation}
\Delta E^{(2)}=\sum_{n\neq 0}{<0|eE_0x_a|n><n|eE_0x_a|0>\over
E_0-E_n}\equiv -{1\over 2}4\pi \alpha_E^aE_0^2
\end{equation}
We find then $\alpha_E^a\sim e^2<x_1^2>/\omega_0$ and
\begin{equation}
\Delta E_{vdw}\sim {\alpha_E^a\alpha_E^b\omega_0\over 4\pi r^6}
\end{equation}
so that it is this {\it self-interaction energy} which is
responsible for the London form.

Casimir and Polder generated a general form for the interaction
potential from quantum mechanics by using two-photon exchange and
fourth-order noncovariant perturbation theory\cite{cp}. Their
result reproduces the simple London form at short
distance---$r\leq {1\over \omega_0}$ but at large distances, when
retardation is important, {\it i.e.}, when a typical quantum
mechanical excitation time $T_{qm}\sim 1/\omega_0$ is smaller than
the time for light to travel between the two particles
$T_\gamma\sim r$ then the London potential, which depends upon the
correlation between the instantaneous positions of the two
systems, breaks down and the interaction evolves into the ong
distance  asymptotic form
\begin{equation}
V_{CasPol}(r)\stackrel{R\rightarrow\infty}\longrightarrow
{-23(\alpha_E^a\alpha_E^b+\beta_M^a\beta_M^b)+
7(\alpha_E^a\beta_M^b+\alpha_E^b\beta_M^a)\over 4\pi
r^7}\label{eq:cas}
\end{equation}
That the very long distance asymptotic form must vary as $1/r^7$ is
clear from simple scaling, as argued by Kaplan\cite{kap}. The
argument is elementary---since polarizabilities have units of
volume, and since the interparticle separation is the only scale in
the problem, the form of the potential must be
$$V\sim -{\alpha_E^a\alpha_E^b\over r^7}.$$  The derivation of the
Casimir-Polder form---Eq. \ref{eq:cas}---within modern quantum field
theory was given by Feinberg and Sucher using dispersive
methods\cite{fs1}.  In an impressive calculation using simple
assumptions involving analyticity they were able to obtain the
Casimir-Polder result.

In this section we shall show how the same form can be obtained in a
much simpler and more direct fashion using the methods of effective
field theory. The basic idea is to calculate the diagram for
two-photon exchange between the two systems and then to retain only
the leading nonanalytic---small momentum transfer---terms, since it
is these pieces which lead to the dominant---large $r$---behavior of
the potential.  We first set the generic framework for our study. We
examine the electromagnetic scattering of two particles---particle
$a$ with mass $m_a$ and incoming four-momentum $p_1$ and particle
$b$ with mass $m_b$ and incoming four-momentum $p_3$. After
undergoing scattering the final four-momenta of particle $a$ is
$p_2=p_1-q$ and that of particle $b$ is $p_4=p_3+q$---{\it cf.}
Figure 1.  Now we need to be more specific.

\begin{figure}
\begin{center}
\epsfig{file=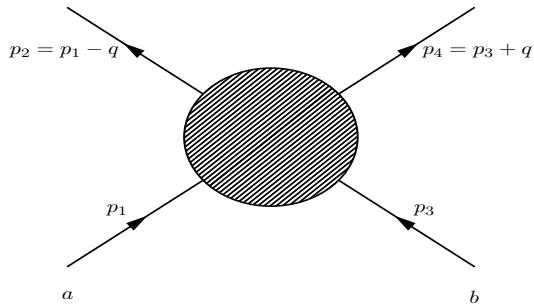,height=4cm,width=7cm} \caption{Basic
kinematics of electromagnetic scattering. }
\end{center}
\end{figure}

\subsection{Spinless Neutral-Spinless Neutral Scattering}

First suppose that the two particles are both neutral and spinless.
Then the leading piece of the electromagnetic amplitude is that for
two photon emission and can be characterized in terms of the
electric and magnetic polarizabilities---$\alpha_E,\beta_M$---which
are in turn defined via the energies\cite{ajp}
\begin{equation}
{}^0\delta E^{(1)}=-{1\over
2}(4\pi\alpha_E\vec{E}^2+4\pi\beta_M\vec{B}^2)\label{eq:pol}
\end{equation}
For a spinless neutral particle $a$ of mass $m_a$ having
four-momentum $p_1$, the amplitude to emit a photon with
polarization $\epsilon_a$ and four-momentum $k$ together with a
second photon having polarization $\epsilon_b$ and four-momentum
$q-k$ is then
\begin{eqnarray}
&&\epsilon_a^{*\alpha}\epsilon_b^{*\beta}{}^0\tau_{\alpha\beta}^a(p_1,k,q-k)\nonumber\\
&=&-i 4\pi \alpha_E^a {1\over m_a^2}(\epsilon_a^{*\alpha} k\cdot
p_1-k^\alpha \epsilon_a^*\cdot p_1 )(\epsilon_{b\alpha}^*(q-k)\cdot
p_1-(q-k)_\alpha \epsilon_b^*\cdot p_1)\nonumber\\
&-&i4\pi \beta_M^a{1\over
m_a^2}(\epsilon^{\alpha\beta\gamma\delta}\epsilon^*_{a\beta}
k_\gamma
p_{1\delta})(\epsilon_{\alpha\rho\sigma\lambda}\epsilon^{*\rho}_b
(q-k)^\sigma p_1^\lambda)
\end{eqnarray}
The two-photon-exchange diagram between spinless neutral particles
is shown in Figure 2 and is of the form
\begin{eqnarray}
&&{}^{00}{\cal M}_{2\gamma}(q)\nonumber\\
&=&{1\over
2!}{(4\pi)^2\over m_a^2m_b^2}\int {d^4k\over
(2\pi)^4}{-i\eta^{\alpha\gamma}\over k^2} {-i\eta^{\beta\delta}\over
(k-q)^2}{}^0\tau^a_{\alpha\beta}(p_1,k,q-k){}^0\tau^b_{s\gamma\delta}(p_3,-k,k-q)\nonumber\\
&=&{1\over 2!}{(4\pi)^2\over m_a^2m_b^2}\int {d^4k\over
(2\pi)^4}{1\over
k^2(k-q)^2}\left[\alpha_E^b\left(\eta^{\alpha\beta}p_3\cdot k
p_3\cdot(k-q)+p_3^\alpha p_3^\beta k\cdot(k-q)\right.\right.\nonumber\\
&-&\left.\left.(k-q)^\alpha p_3^\beta p_3\cdot k-p_3^\alpha
k^\beta p_3\cdot(k-q)\right)+\beta_M^b
(\epsilon^{\lambda\alpha\gamma\delta}k_\gamma p_{3\delta})(
{\epsilon_\lambda}^{\beta\kappa\mu}(k-q)_\kappa p_{3\mu})
\right]\nonumber\\
&\times&\left[\alpha_E^a\left(\eta^{\alpha\beta}p_1\cdot k
p_1\cdot(k-q)+p_{1\alpha} p_{1\beta} k\cdot(k-q)\right.\right.\nonumber\\
&-&\left.\left.(k-q)_\alpha p_{1\beta} p_1\cdot k-p_{1\alpha}
k_\beta p_1\cdot(k-q)\right)+\beta_M^b
({\epsilon^\lambda}_{\alpha\sigma\tau}k^\sigma p_1^\tau)(
{\epsilon_{\lambda\beta\kappa\mu}(k-q)^\kappa p_1^\mu})
\right]\nonumber\\
\quad
\end{eqnarray}

\begin{figure}
\begin{center}
\epsfig{file=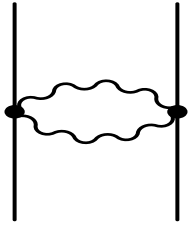,height=3cm,width=2.5cm} \caption{Bubble
diagram used to evaluate the electromagnetic scattering of two
neutral systems. }
\end{center}
\end{figure}

Performing the indicated contractions and integrating, using the
results in Appendix A, we find the result
\begin{equation}
{}^{00}{\cal M}_{2\gamma}(q)=-{Lq^4\over
240}\left[23\left(\alpha_E^a\alpha_E^b+\beta_M^a\beta_M^b\right)
-7\left(\alpha_E^a\beta_M^b+\alpha_E^b\beta_M^a\right)\right]
\end{equation}
where we have defined $L=\log -q^2$.  In order to determine the
potential, we Fourier transform and find, using the results from
Appendix B,
\begin{equation}
{}^{00}V_{2\gamma}(r)=-\int {d^3q\over (2\pi)^3} {\cal
M}_{2\gamma}(q)
e^{-i\vec{q}\cdot\vec{r}}={-23(\alpha_E^a\alpha_E^b+\beta_M^a\beta_M^b)+
7(\alpha_E^a\beta_M^b+\alpha_E^b\beta_M^a)\over 4\pi r^7}
\end{equation}
which is the classic result of Casimir and Polder\cite{cp}.

\subsection{Nonzero Spin Neutral-Spinless Neutral Scattering}

If either neutral particle has spin, the potential becomes more
complex, but is still straightforward.  We must now characterize
the system in terms both of its ordinary electric and magnetic
polarizabilities but also in terms of so-called "spin
polarizabilities."  If the particle $a$ has spin $S_a$, then the
leading order spin-dependent generalization of Eq. \ref{eq:pol}
has the form\cite{dre}
\begin{equation}
{}^{S_a}\delta E_{tot}={}^0\delta
E^{(1)}<S_a,m_{af}|S_a,m_{ai}>+{}^{S_a}\delta E^{(2)}
\end{equation}
where
\begin{eqnarray}
\delta E^{(2)}&=&-4\pi\left[
\gamma_{E1}^a\vec{S}_a\cdot\vec{E}\times\dot{\vec{E}}
+\gamma_{M1}^a\vec{S}_a\cdot\vec{B}\times\dot{\vec{B}}\right.\nonumber\\
&-&\left.2\gamma_{E2}^a\left(\vec{E}\cdot\vec{\nabla}\vec{S}_a\cdot\vec{B}+E_j\vec{S}_a\cdot\vec{\nabla}B_j\right)
+2\gamma_{M2}^a\left(\vec{B}\cdot\vec{\nabla}\vec{S}_a\cdot\vec{E}+B_j\vec{S}_a\cdot\vec{\nabla}E_j\right)
\right]\nonumber\\
\quad\label{eq:sp}
\end{eqnarray}
Here $\vec{S}_a=<S_a,m_{af}|\vec{S}|S_a,m_{ai}>$ and
$\gamma_{E1}^a,\gamma_{M1}^a,\gamma_{E2}^a,\gamma_{M2}^a$ are the
spin-polarizabilities of the particle.  The two-photon vertex of
particle $a$ then has the form
\begin{eqnarray}
&&\epsilon_a^{*\alpha}\epsilon_b^{*\beta}{}^{S_a}\tau_{\alpha\beta}^a(p_1,k,q-k)\nonumber\\
&=&{4\pi\over m_a^2}\left[\alpha_E^a (\epsilon_{a}^{*\alpha} k\cdot
p_1-k^\alpha\epsilon_{a}^*\cdot p_1)(\epsilon_{b\alpha}^*(q-k)\cdot
p_1-(q-k)_\alpha\epsilon_b^*\cdot p)\right.\nonumber\\
&+&\left.\beta_M^a(\epsilon^{\alpha\beta\gamma\delta}\epsilon_{a\beta}^*k_\gamma
p_{1\delta})(\epsilon_{\alpha\rho\sigma\lambda}\epsilon_b^{*\rho}
(q-k)^\sigma p_1^\lambda)\right]<S_a,m_{af}|S_a,m_{ai}>\nonumber\\
&+&\gamma_{E1}^ai\epsilon_{\alpha\beta\gamma\delta}S_a^\alpha\left[(\epsilon_b^{*\beta}
(q-k)\cdot p_1-(q-k)^\beta\epsilon_b^*\cdot
p_1)k^\gamma(\epsilon_a^{*\delta}
k\cdot p_1-k^\delta\epsilon_a^*\cdot p_1)\right.\nonumber\\
&+&\left.(\epsilon_a^{*\beta} k\cdot p_1-k^\beta\epsilon_a^*\cdot
p_1)(q-k)^\gamma(\epsilon_b^{*\delta} (q-k)\cdot
p_1-(q-k)^\delta\epsilon_b^*\cdot p_1)\right]\nonumber\\
&+&\gamma_{M1}^ai\epsilon_{\alpha\beta\gamma\delta}S_b^\alpha\left[
\epsilon^{\beta\rho\lambda\xi}\epsilon_b^{*\rho}(q-k)_\lambda
p_{1\xi}
k_\gamma\epsilon^{\delta\kappa\zeta\sigma}\epsilon_{a\kappa}^*k_\zeta
p_{1\sigma}\right.\nonumber\\
&+&\left.\epsilon^{\beta\kappa\zeta\sigma}\epsilon_{a\kappa}^*k_\zeta
p_{1\sigma}(q-k)^\gamma\epsilon^{\delta\rho\lambda\xi}\epsilon_b^{*\rho}(q-k)_\lambda
p_{1\xi}\right]\nonumber\\
&+&2\gamma_{E2}^a\left[S_b\cdot k(\epsilon_b^{*\rho} (q-k)\cdot
p_1-(q-k)^\rho\epsilon_b^*\cdot
p_1)i\epsilon_{\rho\kappa\zeta\sigma}\epsilon_a^{*\kappa} k^\zeta
p_1^\sigma\right.\nonumber\\
&+&\left.S_b\cdot(q-k)(\epsilon_a^{*\rho} k\cdot
p_1-k^\rho\epsilon_a^*\cdot
p_1)i\epsilon_{\rho\kappa\zeta\sigma}\epsilon_b^{*\kappa}
(q-k)^\zeta p_1^\sigma\right.\nonumber\\
&+&\left.(\epsilon_b^*\cdot k (q-k)\cdot p_1-(q-k)\cdot k
\epsilon_b^*\cdot
p_1)i\epsilon_{\rho\kappa\zeta\sigma}S_a^\rho\epsilon_a^{*\kappa}
k^\zeta p_1^\sigma\right.\nonumber\\
&+&\left.(\epsilon_a^*\cdot (q-k) k\cdot p_1-(q-k)\cdot k
\epsilon_a^*\cdot
p_1)i\epsilon_{\rho\kappa\zeta\sigma}S_a^\rho\epsilon_b^{*\kappa}
(q-k)^\zeta p_1^\sigma\right]\nonumber\\
&+&2\gamma_{M2}^a\left[S_a\cdot k(\epsilon_a^{*\rho} k\cdot
p_1-k^\rho\epsilon_a^*\cdot
p_1)i\epsilon_{\rho\kappa\zeta\sigma}\epsilon_b^{*\kappa}
(q-k)^\zeta p_1^\sigma\right.\nonumber\\
&+&\left.S_a\cdot(q-k)(\epsilon_b^{*\rho} (q-k)\cdot
p_1-(q-k)^\rho\epsilon_b^*\cdot
p_1)i\epsilon_{\rho\kappa\zeta\sigma}\epsilon_a^{*\kappa} k^\zeta
p_1^\sigma\right.\nonumber\\
&+&\left.i\epsilon_{\rho\kappa\zeta\sigma}k^\rho\epsilon_b^{*\kappa}
(q-k)^\zeta p_1^\sigma(\epsilon_a^*\cdot S_a k\cdot p_1-k\cdot S_a
\epsilon_a^*\cdot p_1)\right.\nonumber\\
&+&\left.i\epsilon_{\rho\kappa\zeta\sigma}(q-k)^\rho\epsilon_a^{*\kappa}
k^\zeta p_1^\sigma(\epsilon_b^*\cdot S_a (q-k)\cdot p_1-(q-k)\cdot
S_a \epsilon_b^*\cdot p_1)\right]\label{eq:kl}
\end{eqnarray}
and the scattering amplitude becomes
\begin{equation}
{}^{S_a0}{\cal M}_{2\gamma}(q)={1\over 2!}{(4\pi)^2\over
m_a^2m_b^2}\int {d^4k\over (2\pi)^4}{-i\eta^{\alpha\gamma}\over
k^2} {-i\eta^{\beta\delta}\over
(k-q)^2}{}^{S_a}\tau^a_{\alpha\beta}(p_1,k,q-k){}^0\tau^b_{s\gamma\delta}(p_3,-k,k-q)
\end{equation}
Performing the various contractions and integration, we find
\begin{equation}
{}^{S_a0}{\cal M}_{2\gamma}^{tot}(q)={}^{S_a0}{\cal
M}_{2\gamma}^a(q)+{}^{S_a0}{\cal M}_{2\gamma}^b(q)
\end{equation}
with
\begin{equation}
{}^{S_a0}{\cal M}_{2\gamma}^a(q)=-{Lq^4\over 240
}<S_a,m_{af}|S_a,m_{ai}>\left[23\left(\alpha_E^a\alpha_E^b+\beta_M^a\beta_M^b\right)
-7\left(\alpha_E^a\beta_M^b+\alpha_E^b\beta_M^a\right)\right]
\end{equation}
and
\begin{eqnarray}
{}^{S_a0}{\cal M}_{2\gamma}^b(q)&=&-{Lq^4\over 240 }{i\over
m_a^2}\epsilon_{\alpha\beta\gamma\delta}p_1^\alpha p_3^\beta
q^\gamma S_a^\delta\left[4(\alpha_E^b+\beta_M^b)
(\gamma_{E1}^a+\gamma_{M1}^a)\right.\nonumber\\
&+&\left.20(\alpha_E^b+\beta_M^b)(\gamma_{E2}^a+\gamma_{M2}^a)\right]
\end{eqnarray}
The first piece here is identical to the form found in the spinless
case but is multiplied by the spin-independent factor
$<S_a,m_{af}|S_a,m_{ai}>=\delta_{m_{af}m_{ai}}$.  The second
component, however, is spin-dependent and more interesting. Working
in the center of mass frame with
$\vec{p}_3=-\vec{p}_1\equiv\vec{p}_{CM}$ and taking the
nonrelativistic limit we find
\begin{eqnarray}
{}^{S_a0}{\cal M}_{2\gamma}^b(q)&=&i{Lq^4(m_a+m_b)\over
240m_a^2}\vec{S}_a\cdot\vec{p}_{CM}\times\vec{q}
\left[4(\alpha_E^b+\beta_M^b)(\gamma_{E1}^a+\gamma_{M1}^a)\right.\nonumber\\
&+&\left.20(\alpha_E^b+\beta_M^b)(\gamma_{E2}^a+\gamma_{M2}^a)\right]\nonumber\\
\quad
\end{eqnarray}
Taking the Fourier transform, and noting that
$\vec{r}\times\vec{p}_{CM}=\vec{L}$ is the angular momentum, we
obtain then
\begin{eqnarray}
&&{}^{S_a0}V(r) =-\int{d^3q\over (2\pi)^3}e^{-i\vec{q}\cdot\vec{r}}
{}^{S_a0}{\cal M}_{2\gamma}^{tot}(q)\nonumber\\
&=&<S_a,m_{af}|S_a,m_{ai}>{-23(\alpha_E^a\alpha_E^b+\beta_M^a\beta_M^b)+
7(\alpha_E^a\beta_M^b+\alpha_E^b\beta_M^a)\over 4\pi r^7}\nonumber\\
&+&{m_a+m_b\over m_a^2}
\vec{S}_a\cdot\vec{p}_{CM}\times\vec{\nabla}{1\over \pi
r^7}\left[(\alpha_E^b+\beta_M^b)(\gamma_{E1}^a+\gamma_{M1}^a)
+5(\alpha_E^b+\beta_M^b)(\gamma_{E2}^a+\gamma_{M2}^a)\right]\nonumber\\
&=&<S_a,m_{af}|S_a,m_{ai}>{-23(\alpha_E^a\alpha_E^b+\beta_M^a\beta_M^b)+
7(\alpha_E^a\beta_M^b+\alpha_E^b\beta_M^a)\over 4\pi r^7}\nonumber\\
&+&{m_a+m_b\over m_a^2}\vec{S}_a\cdot\vec{L}{7\over \pi
r^9}\left[(\alpha_E^b+\beta_M^b)(\gamma_{E1}^a+\gamma_{M1}^a)
+5(\alpha_E^b+\beta_M^b)(\gamma_{E2}^a+\gamma_{M2}^a)\right]
\end{eqnarray}
The potential has a spin-independent piece which is simply the
Casimir-Polder result, accompanied by a shorter range spin-orbit
component, which can be identified by its characteristic spin
dependence. Clearly, higher order polarizabilities will lead to new
and shorter range interactions as well as spin-spin correlations in
the case of scattering of two neutral particles both of which carry
spin. However, we will end our discussion here for the
neutral-neutral case and move on the situation that one of the
particles carries a charge.

\section{Spinless Neutral-Charged Particle Interaction}

The long range interaction between a neutral and charged system was
known classically long before its first quantum mechanical
calculation.  In this case the presence of a charge $e$ at the
origin leads to an electric field at location $\vec{r}$ of size
$\vec{E}(\vec{r)}=e\hat{r}/4\pi r^2$.  If there exists a neutral
particle at this location there will be an induced electric dipole
moment $\vec{d}_E=4\pi\alpha_E\vec{E}$.  the corresponding
interaction energy is
$$\delta E=-{1\over 2}\vec{d}_E\cdot\vec{E}(\vec{r})=
-{1\over 2}4\pi\alpha_E\vec{E}^2(\vec{r})=-{\alpha_E \alpha\over
2r^4}$$ where $\alpha=4\pi e^2$ is the fine structure constant.

A full quantum mechanical calculation leads to quantum corrections
to this result and was first performed by Bernabeu and Tarrach using
dispersive methods\cite{bt}.  The problem was later reexamined
dispersively by Feinberg and Sucher\cite{fs2}. The result found for
the leading long range potential between charged and neutral
spinless systems was
\begin{equation}
 V(r)=-{1\over 2}{\alpha\alpha_E\over
 r^4}+{(11\alpha_E+5\beta_M)\alpha\hbar\over 4\pi
 mr^5}+\ldots\label{eq:nv}
\end{equation}
We see that the leading term is classical ($\hbar$-independent) and
agrees with the result found in the simple derivation
above---$V_{cl}(r)\sim -\alpha\alpha_E/r^4$.  However, there exist
additional contributions to the potential which are quantum
mechanical in nature and have the form $V_{qm}(r)\sim
\alpha\alpha_E\hbar/mr^5$.  Numerically these corrections are tiny.
However, such terms are intriguing in that their origin appears to
be associated with zitterbewegung. That is, classically we can
define the potential by measuring the energy when two objects are
separated by distance $r$.  However, in the quantum mechanical case
the distance between two objects is uncertain by an amount of order
the Compton wavelength due to zero point motion---$\delta r\sim
\hbar/ m$. This leads to the replacement
$$V(r)\sim {1\over r^4}\longrightarrow {1\over (r\pm\delta r)^4}
\sim {1\over r^4}\mp 4{\hbar\over mr^5}$$ which is the form found
in our calculations.

\subsection{Spinless Charged--Spinless Neutral Particle}
The EFT evaluation of the charge-neutral interaction proceeds
similarly to that done for two neutral particles, except that the
two photon emission from the charged particle is characterized by
the usual vertices---for a spinless charged particle we have the
one- and two-photon vertices
\begin{eqnarray}
{}^0\tau^{(1)}_\mu(p_2,p_1)&=&-ie(p_1+p_2)_\mu\nonumber\\
{}^0\tau^{(2)}_{\mu\nu}(p_2,p_1)&=&2ie^2\eta_{\mu\nu}
\end{eqnarray}
The relevant diagrams are shown in Figure 3 and the associated
amplitudes are
\begin{eqnarray}
{}^0{\cal M}_{2\gamma}^a(q)&=&e^2{4\pi\over m_b^2}\int {d^4k\over
(2\pi)^4}{\eta_{\alpha\gamma}\eta_{\beta\delta}\over
k^2(k-q)^2}\left[\alpha_E^b\left(\eta^{\alpha\beta}p_3\cdot k
p_3\cdot(k-q)+p_3^\alpha p_3^\beta k\cdot(k-q)\right.\right.\nonumber\\
&-&\left.\left.(k-q)^\alpha p_3^\beta p_3\cdot k-p_3^\alpha
k^\beta p_3\cdot(k-q)\right)+\beta_M^b
(\epsilon^{\lambda\alpha\zeta\sigma}k_\zeta p_{3\sigma})(
\epsilon^{\lambda\beta\kappa\mu}(k-q)_\kappa p_{3\mu})
\right]\nonumber\\
&\times&(2p_1-k-q)^\delta {1\over
(p_1-k)^2-m_a^2}(2p_1-k)^\gamma\nonumber\\
{}^0{\cal M}_{2\gamma}^b(q)&=&2e^2{1\over 2!}{4\pi\over m_b^2}\int
{d^4k\over (2\pi)^4}{\eta_{\alpha\beta}\over
k^2(k-q)^2}\left[\alpha_E^b\left(\eta^{\alpha\beta}p_3\cdot k
p_3\cdot(k-q)+p_3^\alpha p_3^\beta k\cdot(k-q)\right.\right.\nonumber\\
&-&\left.\left.(k-q)^\alpha p_3^\beta p_3\cdot k-p_3^\alpha
k^\beta p_3\cdot(k-q)\right)+\beta_M^b
(\epsilon^{\lambda\alpha\gamma\delta}k_\gamma p_{3\delta})(
{\epsilon_\lambda}^{\beta\kappa\mu}(k-q)_\kappa p_{3\mu})
\right]\nonumber\\
\quad
\end{eqnarray}

\begin{figure}
\begin{center}
\epsfig{file=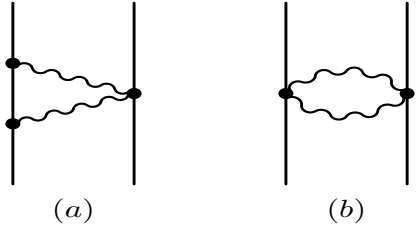,height=3cm,width=5.5cm} \caption{Triangle
and bubble diagrams used to evaluate the electromagnetic scattering
of a charged and a neutral system. }
\end{center}
\end{figure}

Doing the indicated contractions and performing the integration via
the forms given in Appendix A, we find
\begin{eqnarray}
{}^0{\cal M}_{2\gamma}^a(q)&=&-{\alpha q^2\over
4m_a}\alpha_E^b\left(5L+2m_aS\right)\nonumber\\
{}^0{\cal M}_{2\gamma}^b(q)&=&-{\alpha q^2\over
 12m_a}\left(-4\alpha_E^b+5\beta_M^b\right)L
\end{eqnarray}
where we have defined $S=\pi^2/\sqrt{-q^2}$.  Adding, we find
\begin{equation}
{}^0{\cal M}_{2\gamma}^{tot}(q)=-{\alpha q^2\over
12m_a}\left[6m_aS\alpha_E^b+L(11\alpha_E^b+5\beta_M^b)\right]
\end{equation}
whose Fourier transform, using the results given in Appendix B, is
\begin{equation}
{}^0V(r)=-\int{d^3q\over (2\pi)^3}{}^0{\cal M
}_{2\gamma}^{tot}(q)e^{-i\vec{q}\cdot\vec{r}}=-{1\over
2}{\alpha\alpha_E^b\over
 r^4}+{(11\alpha_E^b+5\beta_M^b)\alpha\hbar\over 4\pi m_ar^5}
\end{equation}
in complete agreement with Eq. \ref{eq:nv}.  Now consider the
modifications which result if spin is introduced.

\subsection{Charged Spin 1/2--Spinless Neutral Particle}

In order to see what changes result if the charged particle
carries spin, suppose particle $a$ has spin 1/2.  Then the
calculation goes through as before except that we must use the
one- and two-photon vertices
\begin{eqnarray}
{}^{1\over 2}\tau^{(1)}_\mu(p_2,p_1)&=&-ie\bar{u}(p_2)\gamma_\mu u(p_1)\nonumber\\
{}^{1\over 2}\tau^{(2)}_{\mu\nu}(p_2,p_1)&=&0
\end{eqnarray}
and we find
\begin{eqnarray}
{}^{1\over 2}{\cal M}_{2\gamma}^a(q)&=&-{\alpha q^2\over
12m_a}\left[\alpha_E^b\left(\bar{u}(p_2)u(p_1)(7L+3Sm_a)+{1\over
m_b}\bar{u}(p_2)\not\!{p}_3
u(p_1)(4L+3m_aS)\right)\right.\nonumber\\
&+&\left.\beta_M^b\left(\bar{u}(p_2)u(p_1)(L-3Sm_a)+{1\over
m_b}\bar{u}(p_2)\not\!{p}_3
u(p_1)(4L+3Sm_a)\right)\right]\nonumber\\
{}^{1\over 2}{\cal M}_{2\gamma}^b(q)&=&0
\end{eqnarray}
Using the identity
\begin{equation}
\bar{u}(p_2)\gamma_\mu u(p_1)=\left({1\over 1-{q^2\over
4m_a^2}}\right)\left[{(p_1+p_2)_\mu\over
2m_a}\bar{u}(p_2)u(p_1)-{i\over
m_a^2}\epsilon_{\mu\beta\gamma\delta}q^\beta p_1^\gamma
S_a^\delta\right]\label{eq:id}
\end{equation}
where
$$S_a^\mu={1\over 2}\bar{u}(p_2)\gamma_5\gamma^\mu u(p_1)$$ is the spin vector
and reduces to
$$S_a^\mu\stackrel{NR}{\longrightarrow}(0,\vec{S}_a)=\left(0,\chi_f^{a\dagger}
{1\over 2}\vec{\sigma}\chi_i^a\right)$$
in the nonrelativistic limit, the full amplitude can be written as
\begin{eqnarray}
{}^{1\over 2}{\cal M}_{2\gamma}^{tot}(q)&=&-{\alpha q^2\over
12m_a}\left[
\bar{u}(p_2)u(p_1)\left(\alpha_E^b(11L+6m_aS)+5\beta_M^b\right)\right.\nonumber\\
&+&\left.{i\over
m_a^2m_b}\epsilon_{\alpha\beta\gamma\delta}p_3^\alpha p_1^\beta
q^\gamma
S_a^\delta(4L+3m_aS)(\alpha_E^b+\beta_M^b)\right]\label{eq:oh}
\end{eqnarray}
Taking the nonrelativisitic limit via
\begin{equation}
\bar{u}(p_2)u(p_1)\stackrel{\rm NR}{\longrightarrow}
\chi_f^{a\dagger}\chi_i^a-{i\over
2m_a^2}\vec{S}_a\cdot\vec{p}_2\times\vec{p}_1
\end{equation}
we find the nonrelativistic amplitude in the center of mass frame
\begin{eqnarray}
{}^{1\over 2}{\cal M}_{2\gamma}^{tot}(q)&\simeq&-{\alpha q^2\over
12m_a}\left[\left(6m_aS\alpha_E^b+L(11\alpha_E^b+5\beta_M^b)\right)
\chi_f^{a\dagger}\chi_i^a\right.\nonumber\\
&+&\left.{i\over 2m_a^2}
\vec{S}_a\cdot\vec{p}_2\times\vec{p}_1\left(3{m_a\over
m_b}S(m_a\alpha_E^b+(m_a+m_b)\beta_M^b)\right.\right.\nonumber\\
&+&\left.\left. {1\over
2m_b}L((8m_a-3m_b)\alpha_E^b+(8m_a+3m_b)\beta_M^b\right)\right]
\end{eqnarray}
We observe that the resulting amplitude contains two
components---a spin-independent piece whose form is identical to
that found in the spinless case accompanied by a new
spin-dependent form. Taking the Fourier transform, we find the
effective potential
\begin{eqnarray}
{}^{1\over 2}V(r)&=&\int{d^3q\over (2\pi)^3}{}^{1\over 2}{\cal M
}_{2\gamma}^{tot}(q)e^{-i\vec{q}\cdot\vec{r}}=\left(-{1\over
2}{\alpha\alpha_E^b\over
 r^4}+{(11\alpha_E^b+5\beta_M^b)\alpha\hbar\over 4\pi
 m_ar^5}\right)\chi_f^{a\dagger}\chi_i^a\nonumber\\
 &-&{1\over 2m_a^2}\vec{S}_a\cdot \vec{p}_{CM}\times\vec{\nabla}
 \left({\alpha\over 4m_br^4}\left(m_a\alpha_E^b+(m_a+m_b)\beta_M^b\right)\right.\nonumber\\
 &-&\left.
 {\alpha\hbar\over 8\pi m_bm_br^5}\left((8m_a-3m_b)\alpha_E^b+(8m_a+3m_b)\beta_M^b\right)\right)\nonumber\\
 &=&\left(-{1\over 2}{\alpha\alpha_E^b\over
 r^4}+{(11\alpha_E^b+5\beta_M^b)\alpha\hbar\over 4\pi
 m_ar^5}\right)\chi_f^{a\dagger}\chi_i^a\nonumber\\
 &-&{1\over 2m_a^2m_b}\vec{S}_a\cdot \vec{L}
 \left({\alpha\over r^6}\left(m_a\alpha_E^b+(m_a+m_b)\beta_M^b\right)\right.\nonumber\\
 &-&\left.{5\hbar\alpha\over 8\pi m_ar^7}\left((8m_a-3m_b)\alpha_E^b+(8m_a+3m_b)\beta_M^b\right)\right)
\end{eqnarray}
The potential then has a universal spin-independent form
accompanied by a spin-orbit component, which in turn will be seen
to have a universal structure.  In order to verify this assertion,
we proceed to the case that particle $a$ has unit spin.

\subsection{Charged Spin 1--Spinless Neutral Particle}

In order to verify our conjecture that the spin-orbit piece has a
universal structure, we perform the scattering calculation for the
case of a charged spin 1 particle, which we take to be a $W^+$
boson. In order to determine the correct interaction vertices we
must recall that the electroweak interaction is a gauge theory.
This means that the spin one Lagrangian which contains the
charged-W has the Proca form---
\begin{equation}
{\cal L}=-{1\over 4}(\vec{U}_{\mu\nu})^2+{m^2\over 2}\vec{U}_\mu^2
\end{equation}
but the SU(2) field tensor $\vec{U}_{\mu\nu}$ contains an
additional term on account of the required gauge invariance
\begin{equation}
\vec{U}_{\mu\nu}=\pi_\mu\vec{U}_\nu-\pi_\nu\vec{U}_\mu-
ik\vec{U}_\mu\times\vec{U}_\nu
\end{equation}
where $k$ is the SU(2) electroweak coupling constant.  This
additional term in the field tensor is responsible for the
interactions involving three and four W-bosons and for an "extra"
interaction term which has the form of an anomalous magnetic moment
and, when added to the simple Proca moment, increases the predicted
gyromagnetic ratio from its naive value---$g_{W^\pm}^{\rm
naive}=1$---to its standard model value---$g_{W^\pm}^{\rm
sm}=2$\cite{gau}. The resulting one- and two-photon vertices are
then found to be
\begin{eqnarray}
\tau_\mu(p_2,p_1) &=&-ie\left[(p_2+p_1)_\mu
\epsilon_f^{a*}\cdot\epsilon_i^a-\epsilon_{f\mu}^{a*}\epsilon_i^a\cdot
p_2
-\epsilon_{i\mu}^a\epsilon_f^{a*}\cdot p_1\right.\nonumber\\
&+&\left.\epsilon_{f\mu}^{a*}\epsilon_i^a\cdot(p_1-p_2)-
\epsilon_{i\mu}^a\epsilon_f^{a*}\cdot(p_1-p_2)\right]\nonumber\\
\tau_{\mu\nu}(p_2,p_1)
&=&ie^2(2g_{\mu\nu}\epsilon_f^{a*}\cdot\epsilon_i^a-\epsilon_{i\mu}^a\epsilon_{f\nu}^{a*}
-\epsilon_{f\mu}^{a*}\epsilon_{i\nu}^a)
\end{eqnarray}
where we take the incoming spin 1 particle to have polarization
vector $\epsilon_i^a$ satisfying $\epsilon_i^a\cdot p_1=0$ and the
outgoing particle to have polarization vector $\epsilon_f^a$
satisfying $\epsilon_f^a\cdot p_2=0$.  Evaluating the diagrams shown
in Figure 3 we find then
\begin{eqnarray}
{}^1{\cal M}_{2\gamma}^a(q)&=&{\alpha q^2\over
48m_a}\left[\alpha_E^b\left(2\epsilon_f^{a*}\cdot\epsilon_i^a(29L+12m_aS)
+{1\over m_a^2}\epsilon_f^{a*}\cdot q\epsilon_i^a\cdot
q(20L+9m_aS)\right.\right.\nonumber\\
&-&\left.\left.{2\over m_am_b^2}\epsilon_f^{a*}\cdot
p_3\epsilon_i^a\cdot q(L(m_a-8m_b)-6m_am_bS)-{8\over
m_b^2}\epsilon_f^{a*}\cdot p_3\epsilon_i^a\cdot
p_3L\right.\right.\nonumber\\
&-&\left.\left.{2\over m_am_b^2}\epsilon_f^{a*}\cdot
q\epsilon_i^a\cdot
p_3(L(m_a+8m_b)+6m_am_bS)\right)\right.\nonumber\\
&+&\left.\beta_M^b\left(\epsilon_f^{a*}\cdot\epsilon_i^a8L-{1\over
m_a^2}\epsilon_f^{a*}\cdot q\epsilon_i^a\cdot
q(4L+15m_aS)\right.\right.\nonumber\\
&-&\left.\left.{2\over m_am_b^2}\epsilon_f^{a*}\cdot
p_3\epsilon_i^a\cdot q(L(m_a-8m_b)-6m_am_bS)-{8\over
m_b^2}\epsilon_f^{a*}\cdot p_3\epsilon_i^a\cdot
p_3L\right.\right.\nonumber\\
&-&\left.\left.{2\over m_am_b^2}\epsilon_f^{a*}\cdot
q\epsilon_i^a\cdot
p_3(L(m_a+8m_b)+6m_am_bS)\right)\right]\nonumber\\
{}^1{\cal M}_{2\gamma}^b(q)&=&{\alpha q^2L\over
24m_a}\left[\alpha_E^b\left(-7\epsilon_f^{a*}\cdot\epsilon_i^a+{4\over
m_b^2}\epsilon_f^{a*}\cdot p_3\epsilon_i^a\cdot p_3+{1\over
m_b^2}(\epsilon_f^{a*}\cdot q\epsilon_i^a\cdot
p_3+\epsilon_f^{a*}\cdot
p_3\epsilon_i^a\cdot q)\right)\right.\nonumber\\
&+&\left.\beta_M^b\left(6\epsilon_f^{a*}\cdot\epsilon_i^a+{4\over
m_b^2}\epsilon_f^{a*}\cdot p_3\epsilon_i^a\cdot p_3+{1\over
m_b^2}(\epsilon_f^{a*}\cdot q\epsilon_i^a\cdot
p_3+\epsilon_f^{a*}\cdot p_3\epsilon_i^a\cdot q)\right)\right]
\end{eqnarray}
Summing, we determine the total amplitude
\begin{eqnarray}
{}^1{\cal M}_{2\gamma}^{tot}(q)&=&{\alpha q^2\over
48m_a}\left[\alpha_E^b\left(\epsilon_f^{a*}\cdot\epsilon_i^a4(11L+6m_aS)+{1\over
m_a^2}\epsilon_f^{a*}\cdot q\epsilon_i^a\cdot
q(20L+9m_aS)\right.\right.\nonumber\\
&-&\left.\left.{1\over m_am_b}(\epsilon_f^{a*}\cdot
q\epsilon_i^a\cdot p_3-\epsilon_f^{a*}\cdot p_3\epsilon_i^a\cdot
q)4(4L+3m_aS)\right)\right.\nonumber\\
&+&\left.\beta_M^b\left(\epsilon_f^{a*}\cdot\epsilon_i^a20L-{1\over
m_a^2}\epsilon_f^{a*}\cdot q\epsilon_i^a\cdot
q(4L+15m_aS)\right.\right.\nonumber\\
&-&\left.\left.{1\over m_am_b}(\epsilon_f^{a*}\cdot
q\epsilon_i^a\cdot p_3-\epsilon_f^{a*}\cdot p_3\epsilon_i^a\cdot
q)4(4L+3m_aS)\right)\right]
\end{eqnarray}
In order to make contact with our previous results, we use the
identity
\begin{equation}
\epsilon_{i\mu}^a\epsilon_f^{a*}\cdot q-\epsilon_i^a\cdot
q\epsilon_{f\mu}^{a*}=\left({1\over
4m_a^2-q^2}\right)\left[-4im_a\epsilon_{\mu\beta\gamma\delta}p_1^\beta
q^\gamma S_a^\delta +2(p_1+p_2)_\mu\epsilon_f^{a*}\cdot
q\epsilon_i^a\cdot q\right]\label{eq:kj}
\end{equation}
where we have defined the spin vector
\begin{equation}
S_{a\mu}={-i\over 2m_a}
\epsilon_{\mu\beta\gamma\delta}\epsilon_f^{a*\beta}\epsilon_i^{a\gamma}(p_1+p_2)^\delta
\end{equation}
The amplitude can then be written as
\begin{eqnarray}
{}^1{\cal M}_{2\gamma}^{tot}(q)&=&{\alpha q^2\over
12m_a}\left[\epsilon_f^{a*}\cdot\epsilon_i^a\left(\alpha_E^b(11L+6m_aS)+5\beta_M^bL\right)\right.\nonumber\\
&+&\left.{i\over
m_a^2m_b}\epsilon_{\alpha\beta\gamma\delta}p_3^\alpha p_1^\beta
q^\gamma S_a^\delta(4L+3m_aS)(\alpha_E^b+\beta_M^b)\right.\nonumber\\
&+&\left.{1\over m_a^2}\epsilon_f^{a*}\cdot q\epsilon_i^a\cdot
q\left(\alpha_E^b(4L-3m_aS)-\beta_M^b(20L+27m_aS)\right)\right]
\end{eqnarray}
Comparing with Eq. \ref{eq:oh} we see that both the
spin-independent and dipole terms have a universal form.  There is
an additional quadrupole contribution that presumably is itself
universal if higher spin is considered.

In the nonrelativistic limit we have
\begin{equation}
\epsilon_i^{a0}\simeq -{1\over
m_a}\hat{\epsilon}_i^a\cdot\vec{p_1},\quad\epsilon_f^{a0}\simeq
-{1\over m_a}\hat{\epsilon}_f^a\cdot\vec{p_2}
\end{equation}
so that
\begin{eqnarray}
\epsilon_f^{a*}\cdot\epsilon_i^a&\simeq&
-\hat{\epsilon}_f^{a*}\cdot\hat{\epsilon}_i^a+{1\over
m_a^2}\hat{\epsilon}_f^{a*}\cdot\vec{p}_2\hat{\epsilon}_i^a\cdot\vec{p}_1\nonumber\\
&=&-\hat{\epsilon}_f^{a*}\cdot\hat{\epsilon}_i^a+{1\over
2m_a^2}\hat{\epsilon}_f^{a*}\times\hat{\epsilon}_i^a\cdot\vec{p}_2\times\vec{p}_1
+{1\over
2m_a^2}(\hat{\epsilon}_f^{a*}\cdot\vec{p}_2\hat{\epsilon}_i^a\cdot\vec{p}_1
+\hat{\epsilon}_f^{a*}\cdot\vec{p}_1\hat{\epsilon}_i^a\cdot\vec{p}_2)\nonumber\\
\quad\label{eq:ss}
\end{eqnarray}
Since
\begin{equation}
-i\hat{\epsilon}_f^{a*}\times\hat{\epsilon}_i^a=<1,m_f|\vec{S}|1,m_i>\equiv\vec{S}_a,
\end{equation}
Eq. \ref{eq:ss} becomes
\begin{equation}
\epsilon_f^{a*}\cdot\epsilon_i\simeq
-\hat{\epsilon}_f^{a*}\cdot\hat{\epsilon}_i^a+{i\over
2m_a^2}\vec{S}_a\cdot\vec{p}_2\times\vec{p}_1 +{1\over
2m_a^2}(\hat{\epsilon}_f^{a*}\cdot\vec{p}_2\hat{\epsilon}_i^a\cdot\vec{p}_1
+\hat{\epsilon}_f^{a*}\cdot\vec{p}_1\hat{\epsilon}_i^a\cdot\vec{p}_2)
\end{equation}
Dropping the last term here, which is ${\cal O}(v^2/c^2)$, we find
the nonrelativistic amplitude in the CM frame
\begin{eqnarray}
{}^1{\cal M}_{2\gamma}^{tot}(q)&\simeq&{\alpha q^2\over
12m_a}\left[\left(6m_aS\alpha_E^b+L(11\alpha_E^b+5\beta_M^b)\right)
\hat{\epsilon}_f^{a*}\cdot\hat{\epsilon}_i^a\chi_i\right.\nonumber\\
&+&\left.{i\over 2m_a^2}
\vec{S}_a\cdot\vec{p}_2\times\vec{p}_1\left(3{m_a\over
m_b}S(m_a\alpha_E^b+(m_a+m_b)\beta_M^b)\right.\right.\nonumber\\
&+&\left.\left. {1\over
2m_b}L((8m_a-3m_b)\alpha_E^b+(8m_a+3m_b)\beta_M^b\right)\right.\nonumber\\
&+&\left.{1\over
m_a^2}q:T^a:q\left(\alpha_E^b(4L-3m_aS)-\beta_M^b(20L+27m_aS)\right)\right]
\end{eqnarray}
where
\begin{equation}
q:T^a:q\equiv {1\over 2}(\hat{\epsilon}_f^{a*}\cdot
\vec{q}\hat{\epsilon}_i^a\cdot \vec{q})-{1\over
3}\vec{q}^2\hat{\epsilon}_f^{a*}\cdot\hat{\epsilon}_i^a=
<1,m_f|\vec{S}\cdot\vec{q}\vec{S}\cdot\vec{q}-{2\over
3}\vec{q}^2|1,m_i>
\end{equation}
involves the quadrupole moment.  Taking the Fourier transform we
find the effective potential
\begin{eqnarray}
{}^1V(r)&=&\int{d^3q\over (2\pi)^3}{}^{1\over 2}{\cal M
}_{2\gamma}^{tot}(q)e^{-i\vec{q}\cdot\vec{r}}=\left(-{1\over
2}{\alpha\alpha_E^b\over
 r^4}+{(11\alpha_E^b+5\beta_M^b)\alpha\hbar\over 4\pi
 m_ar^5}\right)\hat{\epsilon}_f^{a*}\cdot\hat{\epsilon}_i^a\nonumber\\
 &-&{1\over 2m_a^2}\vec{S}_a\cdot \vec{p}_{CM}\times\vec{\nabla}
 \left({\alpha\over 4m_br^4}\left(m_a\alpha_E^b+(m_a+m_b)\beta_M^b\right)\right.\nonumber\\
 &-&\left.
 {\alpha\hbar\over 8\pi
 m_bm_br^5}\left((8m_a-3m_b)\alpha_E^b+(8m_a+3m_b)\beta_M^b\right)\right)\nonumber\\
 &+&{1\over m_a^2}\vec{\nabla}:T^a:\vec{\nabla}\left({\alpha\over 4r^4}(\alpha_E^b+9\beta_M^b)
 +{\alpha\hbar\over m_a\pi r^5}(\alpha_E^b-5\beta_M^b)\right)\nonumber\\
 &=&\left(-{1\over 2}{\alpha\alpha_E^b\over
 r^4}+{(11\alpha_E^b+5\beta_M^b)\alpha\hbar\over 4\pi
 m_ar^5}\right)\hat{\epsilon}_f^{a*}\cdot\hat{\epsilon}_i^a\nonumber\\
 &-&{1\over 2m_a^2m_b}\vec{S}_a\cdot \vec{L}
 \left({\alpha\over r^6}\left(m_a\alpha_E^b+(m_a+m_b)\beta_M^b\right)\right.\nonumber\\
 &-&\left.{5\alpha\hbar\over 8\pi
 m_ar^7}((8m_a-3m_b)\alpha_E^b+(8m_a+3m_b)\beta_M^b)\right)\nonumber\\
 &+&{1\over m_a^2}\vec{r}:T^a:\vec{r}\left({24\alpha\over r^8}(\alpha_E^b+9\beta_M^b)
 +{35\alpha\hbar\over m_a\pi r^9}(\alpha_E^b-3\beta_M^b)\right)
\end{eqnarray}
We see then that the potential in the case of spin 0-spin 1
scattering consists of three component. The first is a
spin-independent form which is identical to that found earlier in
the case of spin 0-spin 0 and spin 0-spin 1/2 scattering.  This
piece is accompanied by a shorter range spin-orbit potential
identical to that found in the case of spin 0-spin 1/2 scattering.
Thus both the spin-independent and spin-orbit components are seen to
be universal, in that they have identical forms, independent of
spin. There exists in the case of spin-1 an even shorter range
quadrupole interaction, which we suspect is also universal in
nature.

\subsection{Nonzero Spin Neutral Particle-Spinless Charged Particle}

A fianl possibility is that the charged particle is spinless but the
neutral system carries spin.  In this case, the neutral system is
characterized not only in terms of the electric and magnetic
polarizabilities but also in terms of the four spin polarizabilities
defined in Eq. \ref{eq:sp}  The calculation proceeds as in the case
of a spinless neutral particle, but the two photon vertex Eq.
\ref{eq:kl}   is used.  The resulting diagrams yield

\begin{eqnarray}
{}^0{\cal M}_{2\gamma}^a(q)&=&-{\alpha q^2\over
12m_a}\left[\alpha_E^b\left(15L+6m_aS\right)<S_a,m_{af}|S_a,m_{ai}>\right.\nonumber\\
&+&\left.{i\over
m_am_b}\epsilon_{\alpha\beta\gamma\delta}p_1^\alpha p_3^\beta
q^\gamma
S_a^\delta\left((10L+3m_aS)\gamma_{E1}^b-2L\gamma_{M1}^b\right.\right.\nonumber\\
&+&\left.\left.(26L+9m_aS)\gamma_{E2}^b-(14L+6m_aS)\gamma_{M2}^b\right)\right]\nonumber\\
 {}^0{\cal M}_{2\gamma}^b(q)&=&-{\alpha q^2\over
 12m_a}\left(-4\alpha_E^b+5\beta_M^b\right)L<S_a,m_{af}|S_a,m_{ai}>
\end{eqnarray}
where we have defined $S=\pi^2/\sqrt{-q^2}$.  Adding, we find
\begin{eqnarray}
{}^0{\cal M}_{2\gamma}^{tot}(q)&=&-{\alpha q^2\over
12m_a}\left[\left(6m_aS\alpha_E^b+L(11\alpha_E^b+5\beta_M^b)\right)
<S_a,m_{af}|S_a,m_{ai}>\right.\nonumber\\
&+&\left.{i\over
m_am_b}\epsilon_{\alpha\beta\gamma\delta}p_1^\alpha p_3^\beta
q^\gamma S_a^\delta\left((10L+3m_aS)\gamma_{E1}^b-2L\gamma_{M1}^b\right.\right.\nonumber\\
&+&\left.\left.(26L+9m_aS)\gamma_{E2}^b-(14L+6m_aS)\gamma_{M2}^b\right)\right]
\end{eqnarray}
The effective potential is found as usual by taking the
nonrelativistic limit and Fourier transforming
\begin{eqnarray}
{}^0V(r)&=&\left(-{1\over 2}{\alpha\alpha_E^b\over
 r^4}+{(11\alpha_E^b+5\beta_M^b)\alpha\hbar\over 4\pi
 m_ar^5}\right)<S_a,m_{af}|S_a,m_{ai}>\nonumber\\
&+&{m_a+m_b\over m_am_br^9}\vec{S}_a\cdot\vec{L}
\left((10L+3m_aS)\gamma_{E1}^b-2L\gamma_{M1}^b\right.\nonumber\\
&+&\left.(26L+9m_aS)\gamma_{E2}^b-(14L+6m_aS)\gamma_{M2}^b\right)
\end{eqnarray}

\section{Conclusions}

Above we have examined the long range electromagnetic interaction
between particles with and without spin.  This is not a new
problem---the interaction between two neutral but polarizable
particles was examined in 1948 by Casimir and Polder using old
fashioned perturbation theory\cite{cp}, while that between a neutral
and charged system was treated by Bernabeu and Tarrach in the
mid-1970's using dispersive methods\cite{bt}.  A definitive
dispersive analysis of both problems was given somewhat later by
Feinberg and Sucher\cite{fs1,fs2}.  Here we examined both problems
using ideas from effective field theory and included the
complications associated with spin.  The basic idea of the EFT
approach is that the long range component of the interaction is
generated from the very low momentum transfer region, specifically
from terms which are nonanalytic in $q^2$. One can straightforwardly
isolate such terms from a relativistic Feynman diagram calculation
and the resulting Fourier transform yields the effective potential.
The method is direct and generally {\it much} easier to implement
than that used in earlier treatments.  In this way we have easily
rederived the results of previous authors.  Also, we have included
the effects of spin, which leads to a spin-orbit interaction.  In
the case of a neutral particle, we have used spin polarizabilities
to characterize the structure, while in the case of a charged
particle we have used the usual electromagnetic interaction.  Such
spin-dependent effects are shorter range compared to the leading
spin-independent terms, but they can be identified due to their
characteristic spin dependence. In higher order, if both particles
carry spin then there exists an even shorter-range spin-spin
correlation.  However, we end our discussion here.

\begin{center}
{\bf\large  Appendix A: One loop integration in EFT}
\end{center}

In this section we sketch how our results were obtained.  The
basic idea is to calculate the Feynman diagrams shown in Figure
1a,..e. For simplicity we shall assume spinless scattering. Thus
for Figure 1a we find
\begin{equation}
Amp[1a]={1\over 2!}\int{d^4k\over
(2\pi)^4}{\tau_{\mu\nu}^{(2)}(p_2,p_1)\eta^{\mu\alpha}\eta^{\nu\beta}
\tau_{\alpha\beta}^{(2)}(p_4,p_3)\over k^2(k-q)^2}\label{eqn:a}
\end{equation}
while for Figure 1b
\begin{eqnarray}
Amp[1b]&=&\int{d^4k\over (2\pi)^4}{1\over
k^2(k-q)^2((k-p_1)^2-m_a^2)}\nonumber\\
&\times&\tau^{(2)}_{\mu\nu}(p_4,p_3)\eta^{\mu\alpha}\eta^{\nu\beta}
\tau^{(1)}_\beta(p_2,p_1-k)\tau^{(1)}_\alpha(p_1-k,p_1)\label{eqn:b}
\end{eqnarray}
Here the various vertex functions are listed in section 3, while
for the integrals, all that is needed is the leading nonanalytic
behavior. Thus we use
\begin{eqnarray}
I(q)&=&\int{d^4k\over (2\pi)^4}{1\over k^2(k-q)^2}={-i\over
32\pi^2}(2L+\ldots)\nonumber\\
I_\mu(q)&=&\int{d^4k\over (2\pi)^4}{k_\mu\over k^2(k-q)^2}={i\over
32\pi^2}(q_\mu L+\ldots)\nonumber\\
I_{\mu\nu}(q)&=&\int{d^4k\over (2\pi)^4}{k_\mu k_\nu\over
k^2(k-q)^2}={-i\over 32\pi^2}(q_\mu q_\nu{2\over
3}L-q^2\eta_{\mu\nu}{1\over 6}L +\ldots)\nonumber\\
I_{\mu\nu\alpha}(q)&=&\int{d^4k\over (2\pi)^4}{k_\mu k_\nu
k_\alpha\over k^2(k-q)^2}={i\over 32\pi^2}(-q_\mu q_\nu q_\alpha
{L\over 2}\nonumber\\
&+&(\eta_{\mu\nu}q_\alpha+\eta_{\mu\alpha}q_\nu
+\eta_{\nu\alpha}q_\mu){1\over 12}Lq^2 +\ldots)\nonumber\\
\quad
\end{eqnarray}
with $L=\log -q^2$ for the "bubble" integrals and
\begin{eqnarray}
J(p,q)&=&\int{d^4k\over (2\pi)^4}{1\over
k^2(k-q)^2((k-p)^2-m^2)}={-i\over
32\pi^2m^2}(L+mS)+\ldots\nonumber\\
J_\mu(p,q)&=&\int{d^4k\over (2\pi)^4}{k_\mu\over
k^2(k-q)^2((k-p)^2-m^2)}={i\over
32\pi^2m^2}\nonumber\\
&\times&[p_\mu((1+{1\over 2}{q^2\over m^2})L-{1\over 4}{q^2\over
m}S)-q_\mu(L+{m\over
2}S)+\ldots]\nonumber\\
J_{\mu\nu}(p,q)&=&\int{d^4k\over (2\pi)^4}{k_\mu k_\nu\over
k^2(k-q)^2((k-p)^2-m^2)}={i\over 32\pi^2m^2}\nonumber\\
&\times&[-q_\mu q_\nu(L+{3m\over 8}S)-p_\mu p_\nu{q^2\over
m^2}({1\over 2}L+{m\over 8}S)\nonumber\\
&+&q^2\eta_{\mu\nu}({1\over 4}L+{m\over 8}S)+(q_\mu p_\nu+q_\nu
p_\mu)(({1\over 2}+{1\over 2}{q^2\over m^2})L+{3\over 16}{q^2\over
m S})\nonumber\\
J_{\mu\nu\alpha}(p,q)&=&\displaystyle\int\frac{d^4k}{(2\pi)^4}
\frac{k_\mu k_\nu k_\alpha}{k^2(k-q)^2((k-p)^2-m^2)} \nonumber\\
&=& \frac{-i}{32\pi^2m^2}\bigg[ q_\mu q_\nu
q_\alpha\bigg(L+\frac{5m}{16}S\bigg)+p_\mu p_\nu
p_\alpha\bigg(-\frac{1}{6} \frac{q^2}{m^2}L\bigg) \nonumber\\
\nonumber&+&\big(q_\mu p_\nu p_\alpha + q_\nu p_\mu p_\alpha +
q_\alpha p_\mu p_\nu\big)\bigg(\frac13\frac{q^2}{m^2}L+
\frac1{16}\frac{q^2}{m}S\bigg)\nonumber\\&+&\big(q_\mu q_\nu
p_\alpha + q_\mu q_\alpha p_\nu + q_\nu q_\alpha p_\mu
\big)\bigg(\Big(-\frac13 - \frac12\frac{q^2}{m^2}\Big)L
-\frac{5}{32}\frac{q^2}{m}S\bigg)\nonumber\\
\nonumber &+&\big(\eta_{\mu\nu}p_\alpha + \eta_{\mu\alpha}p_\nu +
\eta_{\nu\alpha}p_\mu\big)\Big(\frac1{12}q^2L\Big)\nonumber\\
\nonumber&+&\big(\eta_{\mu\nu}q_\alpha + \eta_{\mu\alpha}q_\nu +
\eta_{\nu\alpha}q_\mu\big)\Big(-\frac16q^2L -\frac1{16}q^2mS\Big)
\bigg]+\ldots\nonumber\\
\quad
\end{eqnarray}
where $S=\pi^2/\sqrt{-q^2}$ for their "triangle" counterparts.
Similarly higher order forms can be found, either by direct
calculation or by requiring various identities which must be
satisfied when the integrals are contracted with $p^\mu,q^\mu$ or
with $\eta^{\mu\nu}$.

\begin{center}
{\bf\large  Appendix B: Fourier Integrals}
\end{center}

Here we collect the integrals used to calculate the long range
electromagnetic potentials. For the classical effects we use
\begin{eqnarray}
& &\int{d^3q\over
(2\pi)^3}e^{-i\vec{q}\cdot\vec{r}}|\vec{q}|=-{1\over
\pi^2r^4}\nonumber\\
&
&\int{d^3q\over(2\pi)^3}e^{-i\vec{q}\cdot\vec{r}}q_j|\vec{q}|={4ir_j\over
\pi^2r^6}\nonumber\\
& &\int{d^3q\over
(2\pi)^3}e^{-i\vec{q}\cdot\vec{r}}|\vec{q}|^3={12\over
\pi^2r^6}\nonumber\\
&&\int{d^3q\over(2\pi)^3}e^{-i\vec{q}\cdot\vec{r}}q_j|\vec{q}|^3={-i72r_j\over
\pi^2r^8} \label{eq:r1}
\end{eqnarray}
while for the quantum case we utilize
\begin{eqnarray}
& &\int{d^3q\over (2\pi)^3}e^{-i\vec{q}\cdot\vec{r}}\vec{q}^2\log
\vec{q}^2 ={3\over \pi r^5}\nonumber\\
& &  \int{d^3q\over
(2\pi)^3}e^{-i\vec{q}\cdot\vec{r}}{q_j\vec{q}^2\log
\vec{q}^2}={i15r_j\over \pi r^7}\nonumber\\
& & \int{d^3q\over (2\pi)^3}e^{-i\vec{q}\cdot\vec{r}}\vec{q}^4\log
\vec{q}^2 =-{60\over \pi r^7}\nonumber\\
& &  \int{d^3q\over
(2\pi)^3}e^{-i\vec{q}\cdot\vec{r}}{q_j\vec{q}^4\log
\vec{q}^2}={i420r_j\over \pi r^9}\label{eq:r2}
\end{eqnarray}

\begin{center}
{\bf\large Acknowledgement}
\end{center}

This work was supported in part by the National Science Foundation
under award PHY 05-53304.

\end{document}